\documentclass[conference]{IEEEtran}
\IEEEoverridecommandlockouts
\usepackage{cite}
\usepackage{amsmath,amssymb,amsfonts}
\usepackage{algorithmic}
\usepackage{graphicx}
\usepackage{textcomp}
\usepackage{xcolor}
\usepackage{comment}
\usepackage{amssymb}
\usepackage{algorithm}
\usepackage{algorithmic}
\usepackage{url}

\usepackage{listings}
\definecolor{codegreen}{rgb}{0,0.6,0}
\definecolor{codegray}{rgb}{0.5,0.5,0.5}
\definecolor{codepurple}{rgb}{0.58,0,0.82}
\definecolor{backcolour}{rgb}{0.95,0.95,0.92}
\lstdefinestyle{mystyle}{
    backgroundcolor=\color{backcolour},   
    commentstyle=\color{codegreen},
    keywordstyle=\color{magenta},
    numberstyle=\tiny\color{codegray},
    stringstyle=\color{codepurple},
    basicstyle=\ttfamily\footnotesize,
    breakatwhitespace=false,         
    breaklines=true,                 
    captionpos=b,                    
    keepspaces=true,                 
    numbers=left,                    
    numbersep=5pt,                  
    showspaces=false,                
    showstringspaces=false,
    showtabs=false,                  
    tabsize=2
}
\def\BibTeX{{\rm B\kern-.05em{\sc i\kern-.025em b}\kern-.08em
    T\kern-.1667em\lower.7ex\hbox{E}\kern-.125emX}}
\begin{document}

\title{SNOW-SCA: ML-assisted Side-Channel Attack on SNOW-V\\

}

\author{\IEEEauthorblockA{Harshit Saurabh\IEEEauthorrefmark{1},
Anupam Golder\IEEEauthorrefmark{2},
Samarth Shivakumar Titti\IEEEauthorrefmark{1},
Suparna Kundu\IEEEauthorrefmark{3},\\
Chaoyun Li\IEEEauthorrefmark{4},
Angshuman Karmakar\IEEEauthorrefmark{3}\IEEEauthorrefmark{5},
Debayan Das\IEEEauthorrefmark{1}} 
\IEEEauthorblockA{\IEEEauthorrefmark{1}Indian Institute of Science, Bangalore, India} 
\IEEEauthorblockA{\IEEEauthorrefmark{2}Georgia Institute of Technology, USA}
\IEEEauthorblockA{\IEEEauthorrefmark{3}KU Leuven, Belgium}
\IEEEauthorblockA{\IEEEauthorrefmark{4}University of Surrey, UK}
\IEEEauthorblockA{\IEEEauthorrefmark{5}Indian Institute of Technology, Kanpur, India}
\thanks{This work was supported in part by Pratiksha Trust (India), Horizon 2020 ERC Advanced Grant (101020005 Belfort), CyberSecurity Research Flanders with reference number VR20192203, BE QCI: Belgian-QCI (3E230370) (see beqci.eu), and Intel Corporation. 

Angshuman Karmakar is funded by FWO (Research Foundation – Flanders) as a junior post-doctoral fellow (contract number 203056 / 1241722N LV).} 
}

\maketitle
\thispagestyle{plain}
\pagestyle{plain}

\begin{abstract}
This paper presents SNOW-SCA, the first power side-channel analysis (SCA) attack of a 5G mobile communication security standard candidate, SNOW-V, running on a 32-bit ARM Cortex-M4 microcontroller. First, we perform a generic known-key correlation (KKC) analysis to identify the leakage points. Next, a correlation power analysis (CPA) attack is performed, which reduces the attack complexity to two key guesses for each key byte. The correct secret key is then uniquely identified utilizing linear discriminant analysis (LDA). The profiled SCA attack with LDA achieves $100\%$ accuracy after training with $<200$ traces, which means the attack succeeds with just a single trace. Overall, using the \textit{combined CPA and LDA attack} model, the correct secret key byte is recovered with $<50$ traces collected using the ChipWhisperer platform. The entire 256-bit secret key of SNOW-V can be recovered incrementally using the proposed SCA attack. Finally, we suggest low-overhead countermeasures that can be used to prevent these SCA attacks.


\end{abstract}

\vspace{-4mm} 
\section{Introduction}
The evolution of mobile networks, commencing in the late 1970s with the inception of the first Generation (1G) mobile communication technology, has witnessed substantial progress, ultimately leading to the prevalent adoption of the fifth Generation (5G) mobile communication technology. Remarkably, downlink throughput rates have surpassed 1 gigabit per second\cite{caforio_melting_2022}. Each subsequent generation introduced noteworthy advancements; for instance, 2G brought about the introduction of text messaging and encryption, while 3G played a pivotal role in unlocking cyberspace access and enhancing data transfer rates.

In 2018, the $3^{rd}$ Generation Partnership Project (3GPP) tasked the European Telecommunications Standards Institute (ETSI) Security Algorithms Group of Experts (SAGE) with developing new 256-bit cryptosystems for 5G networks~\cite{yang_overview_2020}. These systems must achieve speeds over 20 Gbps on dedicated hardware and general-purpose CPUs, be quantum-safe, and support ultra-reliable low latency communications within a 1ms latency budget. The required key length should also be compatible with the recommendation of the National Institute of Standards and Technology (NIST), which recommends a classical 256-bit security level to provide security against quantum computers. The initiative of 3GPP eventually led to the development of SNOW-V~\cite{3GPPTR33.841-2022}.
\begin{figure}[!t]
\centerline{\includegraphics[scale=0.163]{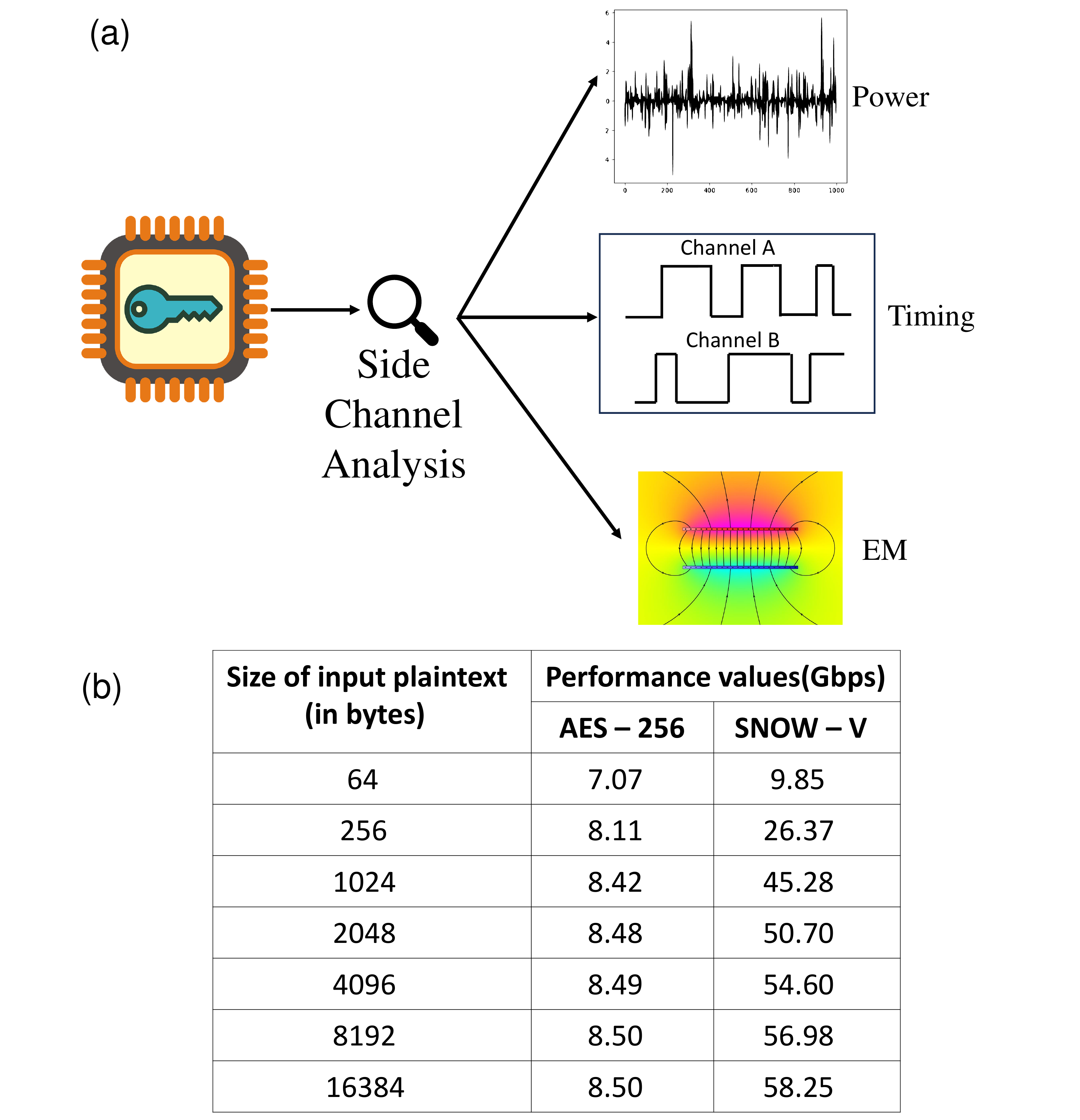}}
\caption{(a) Possible Side channel attacks on a Cryptographic device during encryption  (b) Comparison b/w AES-256 and SNOW-V on performance based on the size of input plaintext \cite{ekdahl_new_2019}.}
\label{intro}
\end{figure}

SNOW-V is a stream cipher proposed by Ekdahl \textit{et~al.}~\cite{ekdahl_new_2019} with the specific goal of being deployed as the \textit{new} encryption primitive in 5G systems. It is closely based on the current 5G standard SNOW 3G, inheriting certain design principles but with modifications suitable for the requirements of 5G networks. SNOW 3G was originally designed as a 128-bit algorithm.  SNOW-V has been designed with 256-bit security in mind, addressing potential vulnerabilities and providing a higher level of cryptographic strength. An ETSI SAGE report~\cite{3GPPSS3-211407} claims that SNOW-V is more resistant to SCA attacks than SNOW 3G, representing a significant security improvement.

Fig.~\ref{intro}(b) shows SNOW-V outperforming AES-256 by $\sim6.5\times$, despite AES-256's optimized assembly code and AES-NI. This improvement in both hardware and software ensures efficient encryption without slowing down high-speed mobile communication.
\subsection{Motivation}
An SCA attack aims to exploit information unintentionally leaked during the execution of cryptographic algorithms. As shown in Fig. \ref{intro}(a), these attacks focus on observing and analyzing various ``side channels" such as power consumption, electromagnetic emission, timing information, etc., about the secret key used in a cryptographic algorithm. 

Block ciphers, such as Advanced Encryption Standard (AES) and Data Encryption Standard (DES), have been extensively studied and applied in secure communication and data storage, while stream ciphers like SNOW-V have not received as much attention in side-channel analysis.

While the rise of post-quantum cryptography (PQC) is a response to the threat posed by quantum computing, stream ciphers, being symmetric-key algorithms, are generally considered less vulnerable to quantum attacks than their asymmetric counterparts as the best-known quantum algorithm \textit{i.e.} Grover search~\cite{grover_search} to break the symmetric-key algorithms gives a quadratic speed-up compared to the classical algorithm. Therefore, for symmetric-key cryptography, the threat of quantum computers can be nullified by doubling the key-length. Nonetheless, the cryptographic community is actively working to ensure that quantum-safe algorithms are available for both symmetric and asymmetric cryptography to maintain the overall security of communication systems in the quantum era. Consequently, as discussed before, to ensure a high level of security for the foreseeable future, 3GPP, in collaboration with ETSI, started a new standardization effort in 2018 for 256-bit symmetric-key algorithms. SNOW-V is one of the candidates for the 5G mobile communication security standard. Hence, it becomes critical to analyze the SCA security of SNOW-V before it is massively deployed in 5G systems. 
\subsection{Contribution}
In this work, we present the first power SCA attack of the stream cipher SNOW-V running on a 32-bit ARM Cortex-M4 microcontroller. In summary, the key contributions of this work are:

\begin{itemize}
  \item We present SNOW-SCA, which is a combined CPA (non-profiled) and ML-based (profiled) attack model for the SNOW-V stream cipher by targeting the update function of the linear feedback shift registers (LFSR). This is the first SCA attack reported on the SNOW-V algorithm (Section III).
  \item We demonstrate and validate successful key recovery using the proposed Known-Initialization vector (IV)-based CPA and linear discriminant analysis (LDA) based attack model on the 32-bit ARM microcontroller (Section IV).
  \item The LDA is used to uniquely identify the correct key byte from the ghost peaks obtained in the CPA attack. The LDA model shows $100\%$ accuracy after training with $< 200$ traces. The minimum traces to disclosure (MTD) for CPA on the measured traces using the Chipwhisperer platform is $< 50$ traces. However, the CPA shows two ghost peaks as the lower significant bit (LSB) could not be uniquely identified. LDA is then used to uniquely identify the correct key using a single trace during the attack phase (Sections III, IV).
  \item Finally, we propose and evaluate different software countermeasures - Boolean masking on the attack points, constant-time implementation of a branching operation to defeat LDA-based profiling, and shuffling. Amongst these, the Boolean masking showed the highest SCA resilience, and the correct key could not be uniquely recovered even after $50,000$ traces, showing $>1000\times$ MTD improvement (Section V).
\end{itemize}

\subsection{Paper Organization}
The paper is structured as follows: Background on SNOW-V is provided in Section II, followed by the presentation of the power SCA attack strategy in Section III. Measurement results of CPA and LDA-based attacks are demonstrated in Section IV, while Section V covers various countermeasures applied to SNOW-V. The paper concludes in Section VI.

\section{Background and Related Works}
\vspace{-1mm}
\subsection{5G Security \& Beyond}
The current cryptographic standards, such as SNOW 3G (designed for 3G mobile networks), encounter new challenges in 5G systems and need adaptation to the evolving technology. Therefore, the shift from SNOW 3G to SNOW-V is crucial, with the increase in demand for security and the establishment of SNOW-V as a standard choice for 5G communications.

SNOW-V has been designed to be compatible and scalable to the upcoming 6G mobile networks. Keeping the futuristic goals in mind, SNOW-V is designed to be PQC-compliant based on the NIST guidelines of 256-bit security for symmetric key algorithms. It is expected that 6G networks will conform to these guidelines to ensure different systems can work together and use secure legacy algorithms that have already been carefully reviewed.


\vspace{-1mm}
\subsection{Why SNOW-V?}
The 5G wireless networks on the horizon are anticipated to deliver high data rates, low latency, and improved Quality of Service. However, with the advent of 5G wireless networks, the demand for security and privacy is expected to be even greater than before.

Serving as a solution to security challenges in previous generations of wireless networks, stream ciphers, such as the 128-bit SNOW 3G, were consistently chosen in 4G and 5G systems. Nevertheless, as 5G networks emerge, algorithms like SNOW 3G find themselves facing new challenges and needing to adapt to the evolving landscape.

To overcome the challenges mentioned earlier, SNOW 3G has recently undergone a revision, evolving into the SNOW-V stream cipher. The SNOW-V stream cipher is a recent addition to the SNOW family of stream ciphers. Its components resemble to those found in other members of the family. The algorithm takes a 256-bit key and a 128-bit initialization vector (IV) as inputs, producing a 128-bit keystream. 
\subsection{Architecture of SNOW-V}
SNOW-V, a synchronous stream cipher, produces a pseudorandom keystream $(z_t)_{t\geq 0}$ with an input of key and initialization vector (IV). At each clock cycle $t$, the message $m_t$ is encrypted by $c_t=m_t\oplus z_t$\ to obtain the ciphertext $c_t$.
Similarly, the decryption is performed by  $m_t=c_t\oplus z_t$.

\begin{figure}[!t]
\centerline{\includegraphics[scale=0.32]{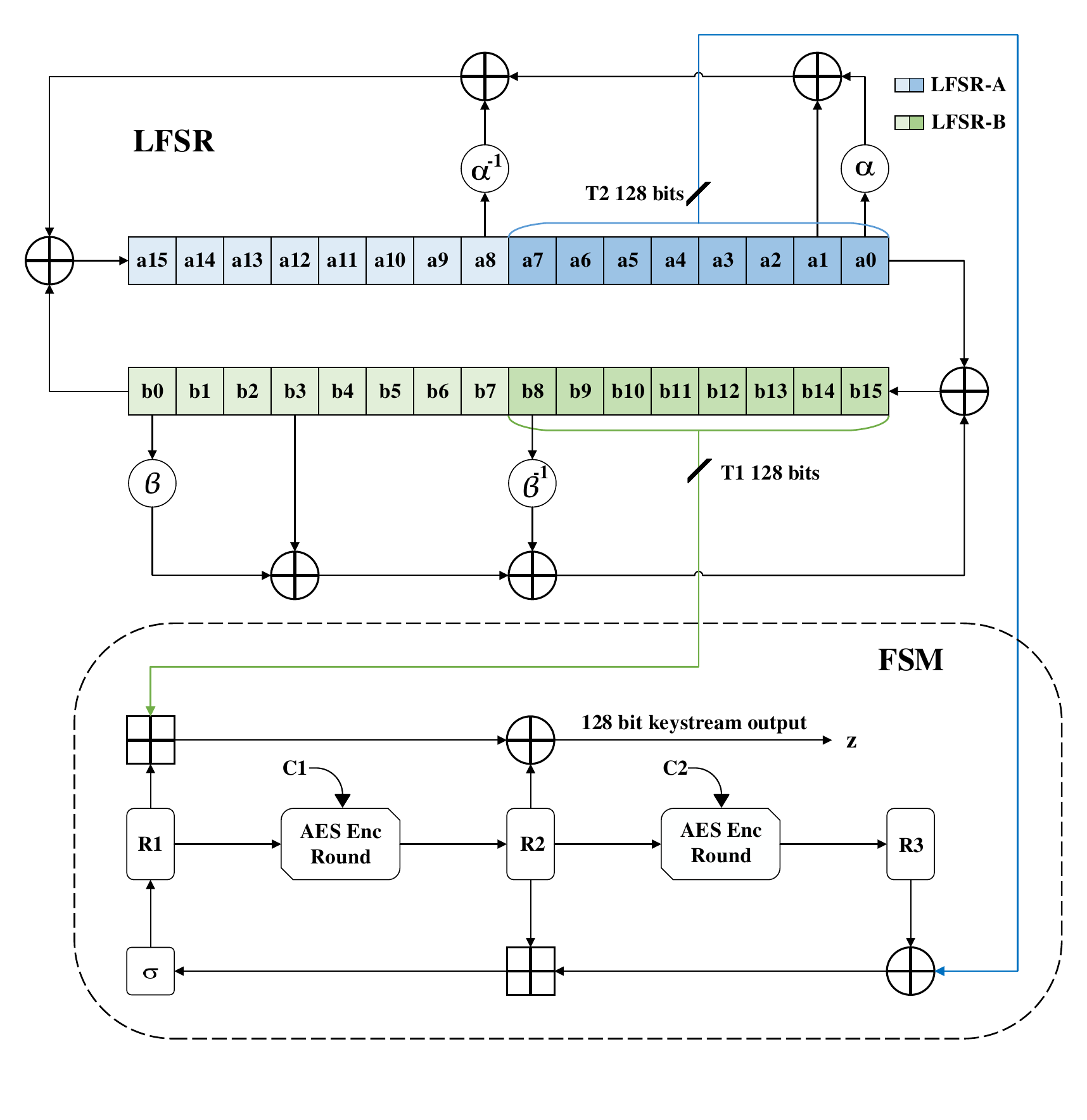}}
\caption{Architecture of SNOW-V, the upper part is the two LFSRs of 512 bits, and the lower part is the FSM consisting of 3 registers and AES round core \cite{ekdahl_new_2019}.}
\label{snowv_arch}
\end{figure}

As shown in Fig.~\ref{snowv_arch}, SNOW-V comprises of primarily two components: an LFSR and a Finite State Machine (FSM). The LFSR part involves two interconnecting shift registers, while the FSM part incorporates two instances of the AES encryption round function and three 128-bit registers.

\subsubsection{{LFSR}}
As shown in Fig. \ref{snowv_arch}, the LFSR part primarily incorporates two 16-stage LFSRs, LFSR-A and LFSR-B, where each stage stores a 16-bit word. In the initialization phase, spanning 16 rounds, the input key and the initialization vector (IV) are fed into and stored in the two LFSRs at their designated positions following the algorithm's specifications. Subsequently, throughout the execution of the algorithm, the contents of the two LFSRs are iteratively updated through a combination of XOR operations and shifts. In each round, these updates occur eight times.

Each 256-bit LFSR consists of 16-bit cells. We denote by $a_0$,....,$a_{15}$ the cells of LFSR-A and analogously $b_0$,....,$b_{15}$ for the cells of LFSR-B.
Each time the LFSR part updates, LFSR-A and LFSR-B clock eight times, i.e., 256 bits of the total 512-bit state of the LFSR part will be updated in a single step, and the two taps T1 and T2 will have fresh values.

\subsubsection{{FSM}}
The FSM  comprises three 128-bit registers, namely R1, R2, and R3 (see Fig. \ref{snowv_arch}). During each cycle, it takes two blocks, T1 and T2, from the LFSR part as inputs and generates a 128-bit keystream block as output. The notation $\boxplus$ denotes parallel addition modulo $2^{32}$ of four 32-bit subwords.
The update logic of the FSM comprises two AES-128 round functions, where the two round keys, C1 and C2, are set to constant values (zero). 
Furthermore, in Fig. \ref{snowv_arch}, $\sigma$ represents a byte-wise permutation of the form:

      $\sigma = [0, 4, 8, 12, 1, 5, 9, 13, 2, 6, 10, 14, 3, 7, 11, 15].$

\subsubsection{{Initialization Phase}}
The initialization phase involves 16 rounds, during which the FSM and the LFSR are updated iteratively. The FSM output $z$ is mixed into LFSR-A in each iteration. Furthermore, during the last two rounds of this initialization phase, the state R1 undergoes an additional update with the key.

\subsubsection{{Keystream Generation Phase}}
In this phase, the LFSRs and the FSM work similarly to the initialization phase, except that the FSM output is not fed into LFSR-A. Thus, the two LFSRs are updated independently of the FSM, while the FSM output makes up the keystream output of the whole algorithm.

\subsection{State-of-the-art SNOW-V Implementation}
Although there have been publications on specific software and hardware implementations of SNOW-V, the clarity regarding its SCA security remains unexplored. SNOW-V is optimized for high-speed software performance, leveraging existing Single Instruction, Multiple Data (SIMD) instructions. Even without AES-NI, SNOW-V can be efficiently implemented using 16/32/64-bit registers. Typically, platforms supporting AES-NI also support other SIMD instructions. Without AES-NI, SNOW-V outperforms AES-256 and even SNOW 3G in terms of speed~\cite{ekdahl_new_2019}. Overall, for large input data, SNOW-V is $6.5\times$ faster than optimized AES-256-CBC for long plaintexts, and even with parallel encryption in AES-256-CTR (instructions interleaving), SNOW-V remains $66\%$ faster~\cite{ekdahl_new_2019}.

Specific hardware implementations of SNOW-V achieve throughput surpassing 20 Gbps. As mentioned in \cite{caforio_melting_2022}, for various libraries like TSMC 90nm and STM 90nm, the throughput reaches 44.91 Gbps, while the throughput for the hardware implementation significantly increases to 628.68 Gbps with NanGate 15nm.
\subsection{Side-Channel Analysis on Stream Ciphers}
While most published works on practical power/EM SCA target block ciphers, there are few works on practical results of power SCA attacks on stream ciphers\cite{kumar_side_2022, strobel_side_nodate}. To successfully attack most block ciphers, focusing on the first or last round is usually enough. However, when analyzing a stream cipher, it is crucial to look at information leaks across several rounds\cite{rechberger_stream_2004}. Also, when examining hardware implementations of stream ciphers, especially those using feedback shift registers, we often combine algebraic attacks with methods from side-channel analysis. 

\section{SNOW-SCA: Attack Methodology}
In this section, we will describe our SNOW-SCA attack methodology on SNOW-V.

\vspace{-2mm}
\subsection{Attack Steps}
\begin{figure}[!t]
\centerline{\includegraphics[scale=0.57]{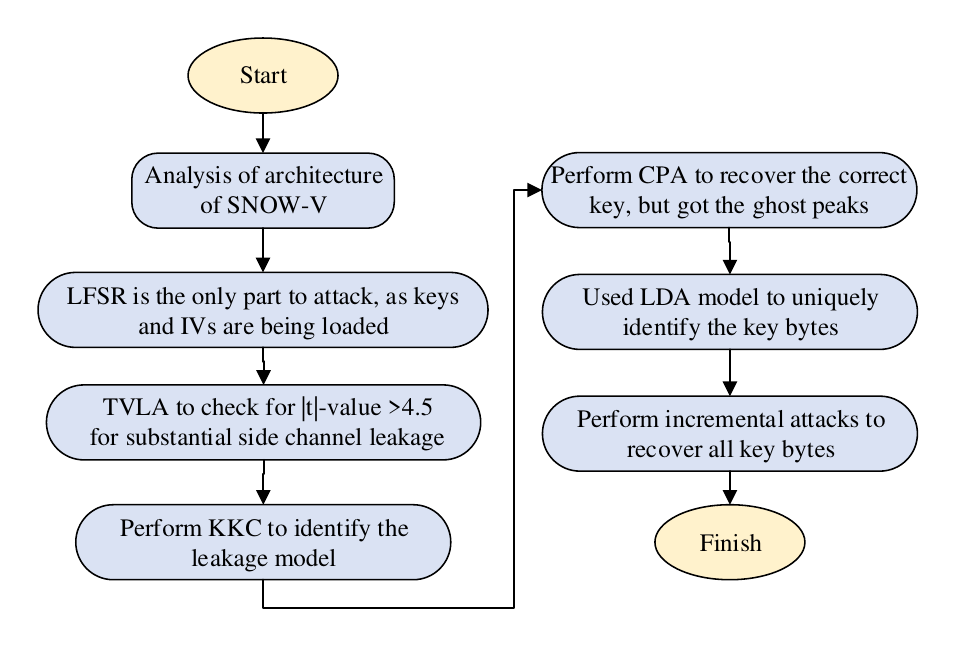}}
\caption{Flowchart for the SNOW-SCA attack methodology}
\vspace{-4mm}
\label{flowchart_attack_model}
\end{figure}

Fig.~\ref{flowchart_attack_model} highlights the attack steps leading to the full key recovery in SNOW-V. Our initial investigation focuses on analyzing the architecture of SNOW-V, revealing that the LFSR is the most vulnerable point of attack. This is due to LFSR's containment of keys and IVs during the initialization phase (Section III-B). We use Welch's t-test hypothesis to check for the time points corresponding to $|t|$-values $> 4.5$, indicating any data-dependent side-channel leakage (Section III-C).

Once the potential point of attack is identified, we perform a Known-Key Correlation (KKC) analysis to verify and validate our attack model (Section III-D). The objective is to identify and analyze the leakage patterns associated with the chosen point of attack. Subsequently, CPA \cite{brier_correlation_2004} is utilized to perform the attack targeting one key byte at a time, ultimately leading to the recovery of the correct key (Section III-E). The results from the CPA showed the presence of some ghost peaks associated with incorrect keys. To address this, we employed a Linear Discriminant Analysis (LDA) model to predict the Least Significant Bit (LSB) of the targeted key byte and uniquely identify the correct key byte (Section III-F). Additionally, we demonstrate how an incremental attack can be performed to recover all key bytes of SNOW-V (Section III-G).
\vspace{-1mm}
\subsection{Analysis of the LFSR}\label{sec:analysis_LFSR}
The primary focus of the proposed SNOW-SCA attack model is the LFSR in the SNOW-V architecture. During the initialization phase, all key bytes are allocated to the LFSR. In the FSM part, the values of the two round key constants, $C_1$ and $C_2$, are set to zero. Consequently, there is no rationale for targeting the AES part, as its purpose was only to randomize the sequence.

According to the initialization phase mentioned in the specification \cite{ekdahl_new_2019},

\begin{equation*}
    \begin{array}{rl}
    (a_{15}, a_{14},......, a_8) & \leftarrow (k_7, k_6,......, k_0) \\
    (a_7, a_6,......, a_0) &  \leftarrow (iv_7, iv_6,......, iv_0) \\
    (b_{15}, b_{14},......, b_8)  &  \leftarrow (k_{15}, k_{14},......, k_8) \\
    (b_7, b_6,......, b_0) & \leftarrow (0, 0,......, 0)
    \end{array}
\end{equation*} 
where the \textit{secret key  K} = $(k_{15}, k_{14},......., k_1, k_0)$, the IV = $(iv_7, iv_6,..., iv_1, iv_0)$, and each of $k_i,iv_j$, $0 \leq i \leq 15, 0 \leq j \leq 7$, is a 16-bit vector.

These equations highlight that the primary target for the attack can be the LFSR, particularly where the key is used. The provided C-code in the paper \cite{ekdahl_new_2019} provides the SNOW-V implementation details and is used for our proposed attack.

\begin{lstlisting}[caption=$lfsr\_update()$ function, label={lst:unmasked_lfsr}, language=C]
void lfsr_update(void)
{
   for (int i = 0; i < 8; i++)
{ 
   u16 u = mul_x(A[0], 0x990f) ^ A[1] ^ mul_x_inv(A[8], 0xcc87) ^ B[0];
   u16 v = mul_x(B[0], 0xc963) ^ B[3] ^ mul_x_inv(B[8], 0xe4b1) ^ A[0];
   for (int j = 0; j < 15; j++)
   { 
   A[j] = A[j + 1];
   B[j] = B[j + 1];
   }
   A[15] = u;
   B[15] = v;
}}
\end{lstlisting}

If we dive more into the C-code (Listing \ref{lst:unmasked_lfsr}) and analyze the $lfsr\_update()$ function, the parameter $u$ is defined as:
\vspace{-2mm}
\begin{multline}
    u = mul\_x(A[0], 0x990f)\oplus A[1] \\ \oplus\,mul\_x\_inv (A[8], 0xcc87) \oplus B[0]
    \vspace{-2mm}  
\end{multline}

\begin{figure*}[!t]
\centerline{\includegraphics[scale=0.3]{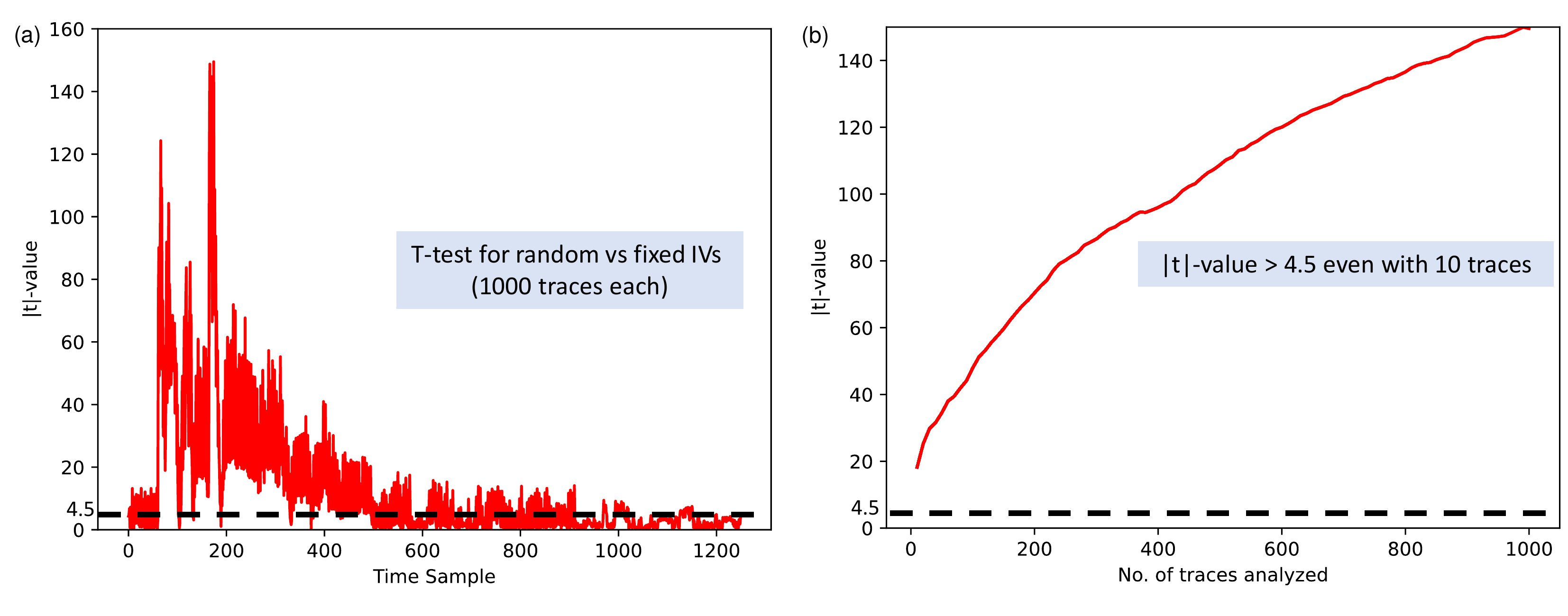}}
\vspace{-4mm}
\caption{Fixed-vs-random TVLA on the measured SNOW-V traces: (a) TVLA for 1K traces across time samples (b) Incremental TVLA showing that the $|t|$-value cross the threshold of 4.5 with $<10$ traces. }
\vspace{-4mm}
\label{TVLA}
\end{figure*}

The above equation is a function of $A[8]$, which contains the first two bytes of the secret key. Therefore, considering the first iteration ($i=0$, refer to Listing \ref{lst:unmasked_lfsr}) for the first key byte, $u$ can be expressed as a function of $f(A[0], A[1], A[8])$, where $B[0]$ can be neglected as it is initialized with all zeros in the initial iteration. 
Note that A[8] contains 16 bits of the key, and we attack 8 bits at a time. Unless otherwise mentioned, most analysis throughout this paper is shown on the $1^{st}$ key byte of SNOW-V, which is $A[8][7:0]$ (lower 8 bits of A[8]), henceforth referred to as A[8] in this paper.

As shown in Fig. \ref{snowv_arch}, LFSR-A contains sixteen 16-bit cells. By analyzing the equation $u$ for the first iteration ($i=0$), the first key byte A[8][7:0] can be recovered. For the $1^{st}$ iteration, $u$ is a function of $f(A[0], A[1], A[8])$. In general, the equation for $u$ is shown in Eqn. 1. By analyzing the values of $u$ and $v$ across the eight iterations, all 32 key bytes can be recovered progressively (discussed in detail in Section III-G).

\begin{lstlisting}[caption=$mul\_x\_inv()$ code, label={lst:mul_x_inv}, language=C]
u16 mul_x_inv(u16 v, u16 d)
{ if (v & 0x0001) return(v >> 1) ^ d;
else return (v >> 1);
}
\end{lstlisting}

From Listing~\ref{lst:unmasked_lfsr} and Listing~\ref{lst:mul_x_inv}, it is evident that for $u$ computation, the information about the LSB of the key byte under attack A[8] ($A[8][0]$) is lost. This is due to the 1-bit right shift within the $mul\_x\_inv()$ function. This has consequences during the CPA attack, resulting in multiple ghost peaks, which we will discuss in Section III-E.


\vspace{-1mm}
\subsection{TVLA of SNOW-V}
Test Vector Leakage Assessment (TVLA), or the statistical t-test, determines if any data-dependent variation exists between the two sets of traces - fixed and random (fixed-vs-random IV, in the case of SNOW-SCA). 
TVLA on SNOW-V was performed for fixed vs random set of 1K traces as shown in Fig.~\ref{TVLA}(a), across different time samples. In the case of fixed traces, the IV remained constant, while for random traces, the IV varied randomly. The criterion for significant data-dependent leakage is based on the $|t|$-value crossing the threshold of 4.5. The results of the TVLA indicate that there is substantial side-channel leakage, as it crosses $|t|$-values surpassing the 4.5 threshold. Fig.~\ref{TVLA}(b) illustrates the execution of TVLA across varying numbers of traces, with the maximum $|t|$-value taken for each trace. Notably, the TVLA plot indicates that even with just ten traces, the $|t|$-value surpasses the threshold of 4.5.

\vspace{-1mm}
\subsection{Known-Key Correlation (KKC) analysis}
In SCA, KKC analysis correlates variations in side-channel information with a known key, facilitating the detection of patterns or leakage that can be exploited to uncover information about the secret key employed in a cryptographic algorithm.

\begin{figure*}[!t]
\centerline{\includegraphics[scale=0.3]{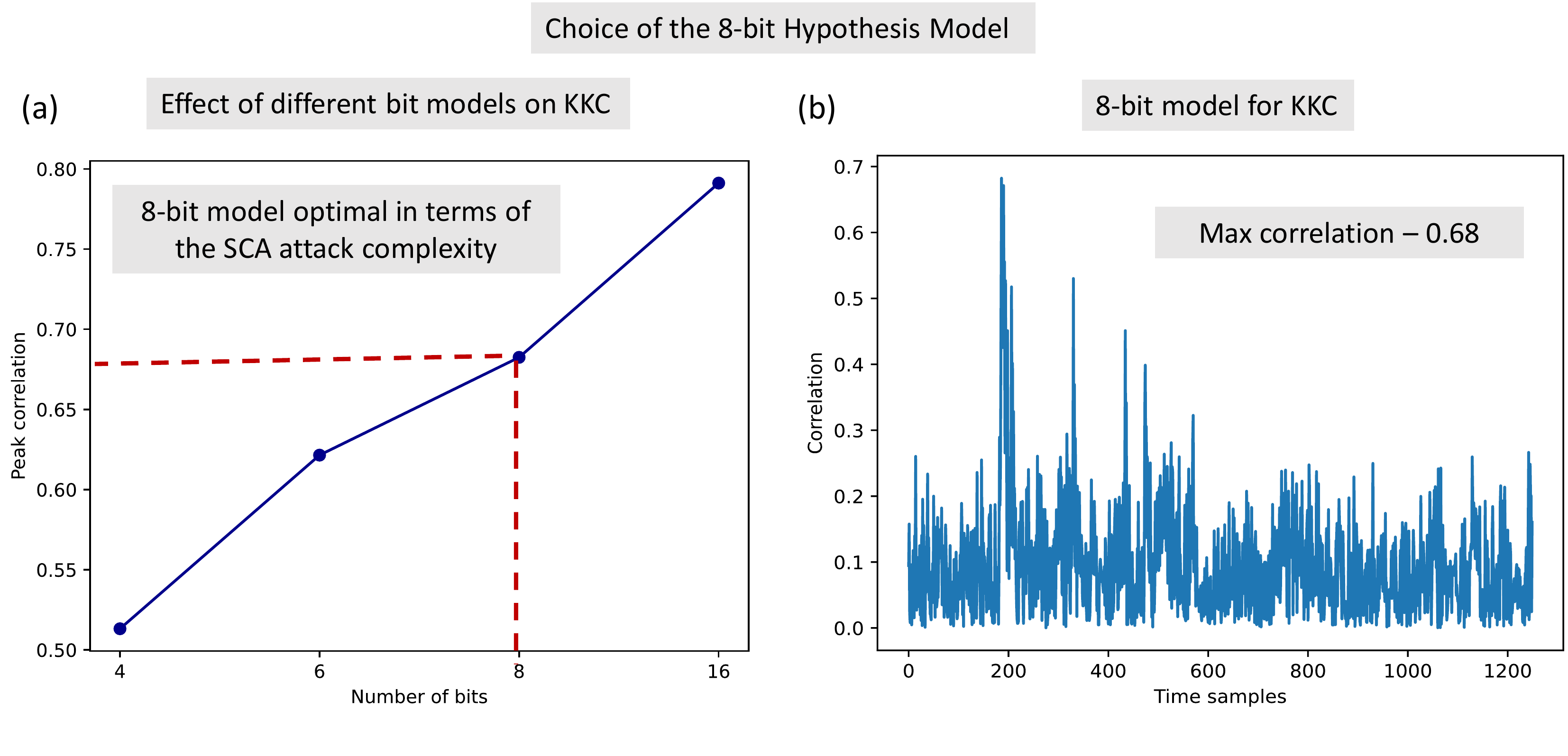}}
\vspace{-3mm}
\caption{(a) Analyzing the 4-bit, 6-bit, 8-bit, and 16-bit models for comparison (b) Considering the measured 8-bit model because it reduces the complexity from $2^{16}$ to $2^8$ and exhibits noticeable correlation, particularly when compared to the 4-bit and 6-bit models. }
\label{kkc_model}
\end{figure*}

\begin{figure*}[!t]
\centerline{\includegraphics[scale=0.3]{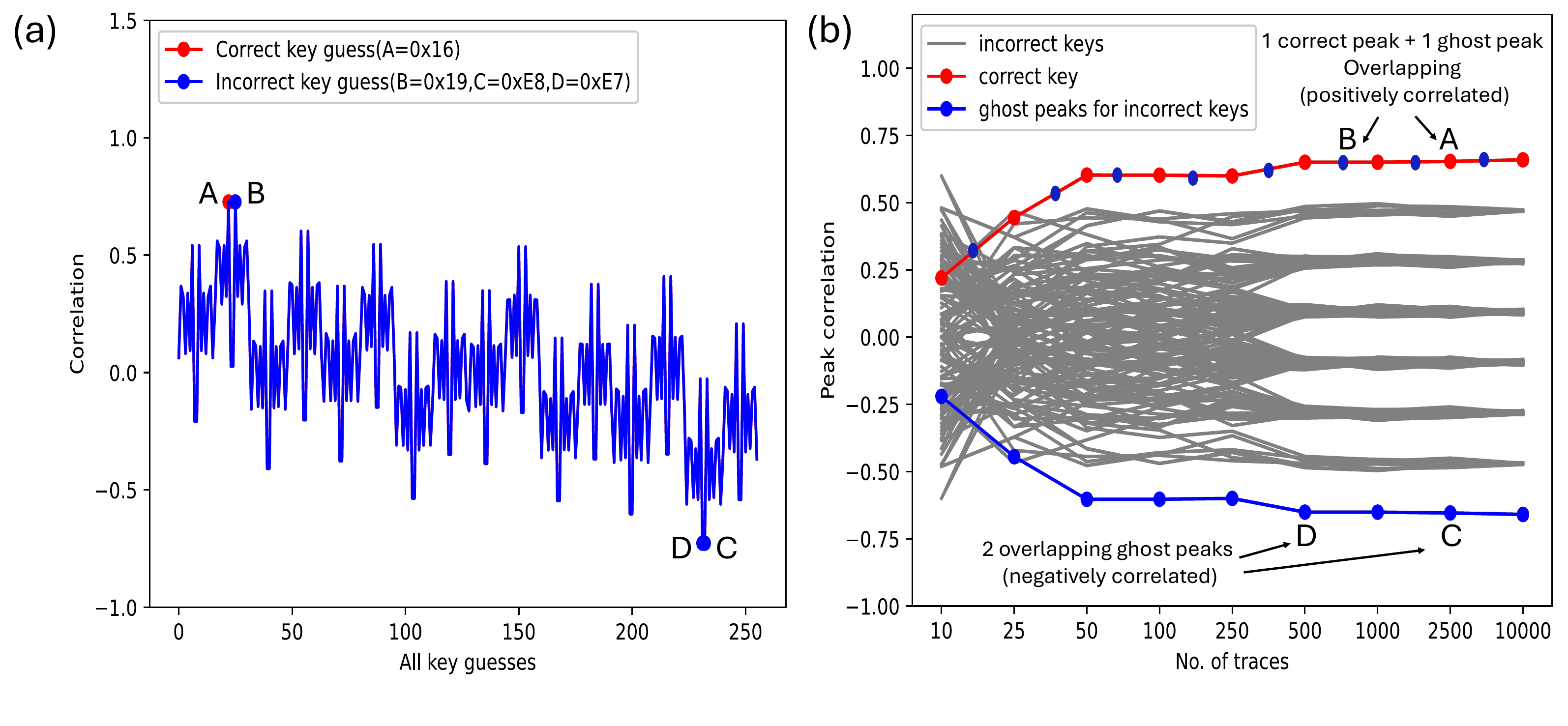}}
\caption{(a) Simulated CPA on SNOW-V (b) MTD plot showing the correct key separates after $\sim30$ traces.}
\vspace{-4mm}
\label{cpa+mtd}
\end{figure*}

The correlation plot in Fig. \ref{kkc_model}(b) illustrates that the peak corresponds to the leakage point of the $u$ operation for the $1^{st}$ key byte. Similarly, when we repeat for multiple keys, distinct peaks are observable at different time samples.
KKC is just an intermediate step to determine the leakage point of the model. For the KKC analysis, we fixed the key and varied the IV to determine the Hamming Weight (HW) of all 16 bits of $u$. Subsequently, we transitioned from a 16-bit to an 8-bit model to deduce the initial byte of the key. The rationale for opting for the \textit{8-bit model} lies in its ability to reduce the attack complexity from $2^{16}$ to $2^8$. While a 2-bit or 4-bit model could have been employed, these would have suffered from a lower signal-to-noise ratio (SNR). Hence, we chose the 8-bit model for its optimal attack complexity (Fig. \ref{kkc_model}(a)). 

While employing the 8-bit model to compute the $u$ hypothesis, it is noteworthy that the HW for the hypothesis is specifically calculated for the \textit{7 bits of the $u$ hypothesis}. This stems from the fact that the LSB of the key byte is excluded throughout the entire analysis. Upon closer examination of the equation for $u$ as outlined in Section \ref{sec:analysis_LFSR}, it becomes evident that the key situated in A[8] of the $mul\_x\_inv()$ function undergoes a right shift by one bit (refer to Listing \ref{lst:mul_x_inv}).

\vspace{-1mm}
\subsection{CPA on SNOW-V}
\vspace{-0.5mm}
Following the KKC analysis, we perform CPA targeting the $u$ and $v$ operations in the $lfsr\_update()$ function to extract the secret key byte. Before we move to the measurement results with CPA, we first perform CPA on simulated traces, which makes the analysis faster.

\begin{figure}[!t]
\centerline{\includegraphics[scale=0.19]{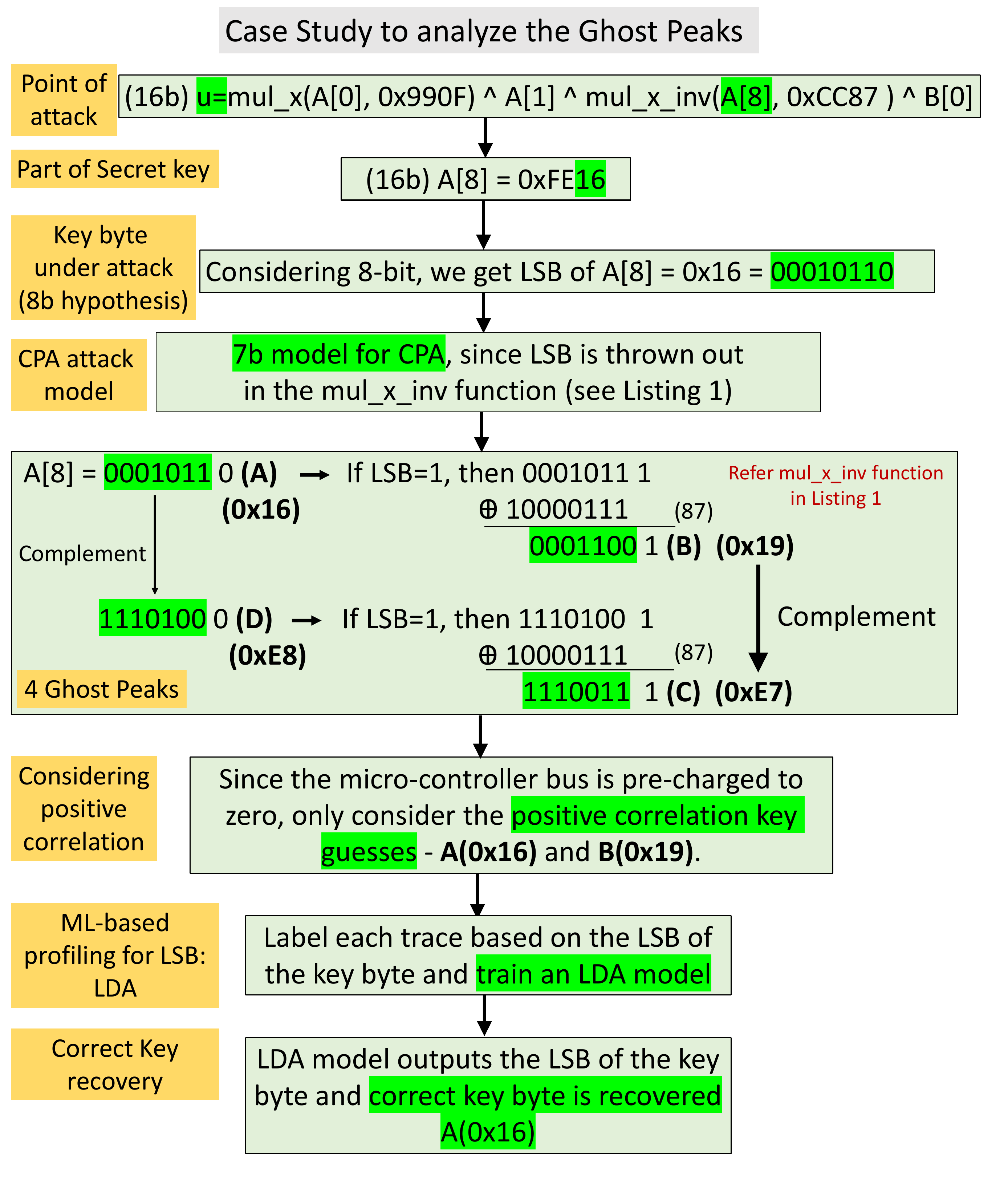}}
\caption{
A Case Study demonstration to analyze and understand the ghost peaks and how the correct key byte can be uniquely recovered using the proposed combined CPA+LDA attack model.}
\label{cpa-graph}
\end{figure}

For the simulated CPA (Fig.~\ref{cpa+mtd}(a)), we first compute the 16-bit $u$ using the LFSR update function and determine its HW. The HW of $u$ serves as the representation for the simulated traces. Now, for our 8-bit key hypothesis, the key byte under attack A[8] is varied from 0 to 255, and the corresponding hypothetical 8-bit $u$ is computed. As discussed in the previous sub-section, although our key guess is 8-bit, we applied the HW to the 7-bit of $u$ (8-bit key hypothesis, but 7-bit attack model) due to the shift in the LSB of A[8] (refer to Fig. \ref{cpa-graph}). 

As shown in Fig.~\ref{cpa+mtd}(a), the CPA attack reveals four key guesses having the highest correlation - two positive peaks (A, B) and two negative peaks (C, D). Now, out of these four peaks, one is the correct key, and the other three are ghost peaks. For the 32-bit ARM Cortex-M4 microcontroller, the data bus is pre-charged to zero, and hence, the CPA should show positive peaks for the correct key byte. Hence, we can discard the two negative ghost peaks (C, D). However, we still cannot distinguish the correct key byte between A and B. 

Fig.~\ref{cpa-graph} shows a case study to analyze the ghost peaks observed with CPA. As shown in Fig.~\ref{cpa-graph}, the four possibilities for the top key guess are attributed to the shift in the LSB of the key byte under attack due to the $mul\_x\_inv()$ function (refer to Listing \ref{lst:mul_x_inv}). Hence, the remaining 7 bits of the key byte contribute to the power SCA leakage correlated to the HW. Since the $u$ and $v$ operations are fully linear (part of LFSR), the complement of the 7-bit value also correlates negatively, showing the ghost peaks. In this case study (Fig. \ref{cpa-graph}), the original key byte under attack ($A[8]$) is 0x16 (A). Now, for this key byte value, the LSB is 0, which is thrown out due to the $mul\_x\_inv()$ function. The other possibility that correlates to the 7-bit attack model is the case when LSB is 1. In that case, the ghost key byte will be the 7-bit XOR of the $0$x$16 >> 1$ (0x0B) and 0x07. The LSB being one would result in the ghost key byte as 0x19 (B). Similarly, we would obtain negative correlation peaks for the complement of the upper 7 bits of A and B (LSB remaining the same), resulting in 0xE8 (C) and 0xE7 (D) as two other ghost peaks. In summary, the LSB bit being 0 or 1, and the complement of the remaining 7 bits create the four combinations (A, B, C, D) as observed in Fig.~\ref{cpa+mtd}(a, b).

The MTD plot in Fig.~\ref{cpa+mtd}(b) shows that the correct key byte (along with the ghost peaks) can be recovered in $\sim30$ traces for the simulated CPA. The measured  CPA results are discussed in Section IV-B.



\vspace{-2mm}
\subsection{Linear Discriminant Analysis (LDA) Model}\label{sec:LDA_model}
\vspace{-0.5mm}
Following the CPA attack, we have two positive correlation peaks, of which one is the correct key. The ghost peak appears due to the unknown LSB, as discussed earlier. The LSB gets thrown out due to the $mul\_x\_inv()$ function (refer to Listing \ref{lst:mul_x_inv}). However, the LSB affects the timing of the $mul\_x\_inv()$. If the LSB of the key byte A[8] is 1, then a right shift and an XOR operation are performed; otherwise, if the LSB is 0, only the shift is performed. 

To model this leakage, we utilize the Linear Discriminant Analysis (LDA) by modeling each trace based on the LSB of the key byte A[8] under attack. In recent SCA attacks, LDA has been used successfully in profiled SCA \cite{choudary_efficient_2014, 10.1145/3465380}. LDA is a machine-learning (ML)-based dimensionality reduction technique that maximizes inter-class separation. We utilize LDA in this work as a binary classifier.

\begin{figure}[!t]
\centerline{\includegraphics[scale=0.5]{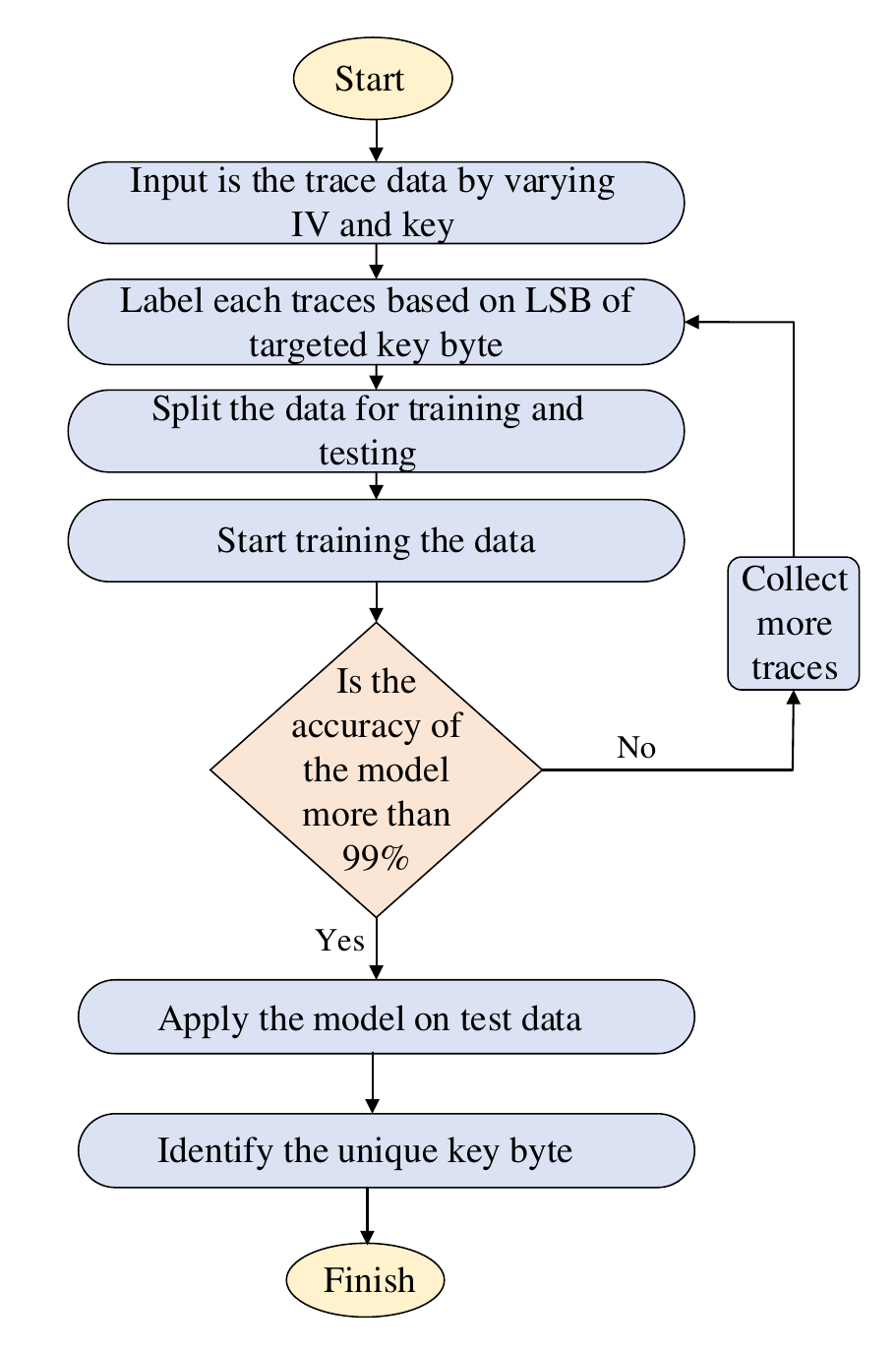}}
\caption{Flowchart for the LDA model used to identify the key bytes uniquely.}
\label{LDA flowchart}
\end{figure}


The LDA model is trained to learn the LSB of the key byte under attack from the power trace. During the training phase, the goal is to achieve $100\%$ accuracy so that during the attack/test phase, the correct LSB is identified with a single trace. This would eliminate the false peak (B) after CPA and help uniquely identify the correct secret key byte (A) (refer Fig. \ref{cpa+mtd}). The LDA measurement results are demonstrated in Section IV-B.

\vspace{-2mm}
\subsection{Incremental Attack to Recover All Key Bytes}
Through an incremental attack, we can recover all correct key guesses. Algorithm 1 demonstrates the steps that allow for retrieving all potential key guesses. From the algorithm, it's clear that, after the fourth iteration, i.e., for $i\geq5$, there will be XOR-ing between two 16-bit words of the same keys, but one of the keys will be known to us in the previous attack steps, i.e., during $i\leq4$. So, XOR-ing between the two 16-bit words of the same key with one known value will give us another value.

\begin{figure}[!t]
\vspace{-4mm}
\centerline{\includegraphics[scale=0.55]{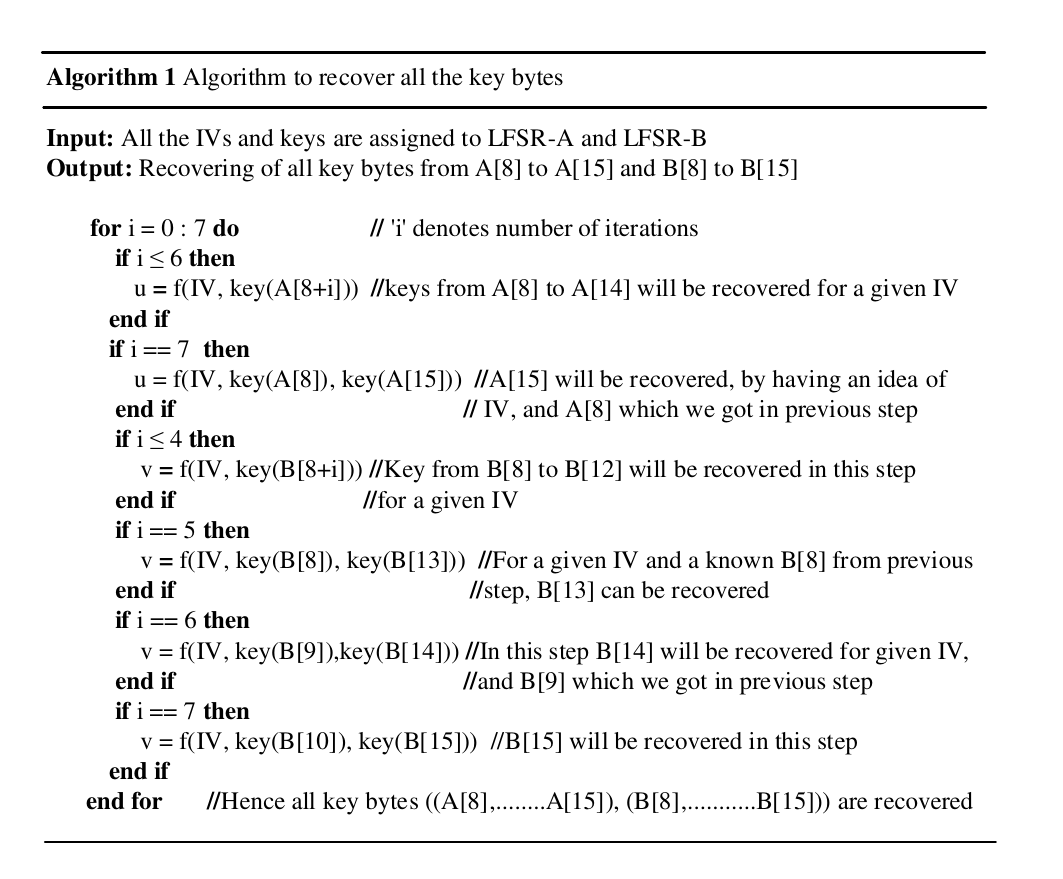}}
\label{algorithm for all key byte}
\vspace{-4mm}
\end{figure}

In the first iteration of the LFSR update function ($i=0$) (refer to Listing \ref{lst:unmasked_lfsr}), the equations for $u$ and $v$ involve performing XOR operations on the IV and the key, located in $A[8]$ and $B[8]$, respectively. Likewise, for $i=1,2,3,4$, the equations for $u$ and $v$ entail XOR operations involving the IV and keys situated in arrays$ (A[9], A[10], A[11], A[12])$ and $(B[9], B[10], B[11], B[12])$, respectively. 

Advancing to the subsequent iteration $i=5$, the equation for $u$ involves XOR-ing the IV with the key found in $A[13]$. Meanwhile, the $v$ equation XORs the IV with two 16-bit words of the same key, one from $B[8]$ (which is known from the first iteration at $i=0$) and the other from $B[13]$. The only unknown key in this iteration is the one in $B[13]$, making it recoverable during this specific iteration.

Likewise, in the case of $i=6$, the equation for $u$ involves a simple XOR operation between the IV and the key in $A[14]$. However, the $v$ equation XORs the IV with two 16-bit words of the same key, one from $B[9]$ (known from the second iteration, $i=1$) and the other from $B[14]$. So, the unknown key from $B[14]$ can be recovered.

In the final iteration ($i=7$), the equation for $u$ involves XOR-ing the IV with two 16-bit words of the same key from $A[8]$ (known from the first iteration, $i=0$) and $A[15]$. Since $A[8]$ is already known, this iteration allows us to determine the previously unknown key $A[15]$. Similarly, for the $v$ equation, XOR operations are performed between the IV and two 16-bit words of the same key from $B[10]$ and $B[15]$. With $B[10]$ being known from the third iteration ($i=2$), we can now determine the previously unknown key $B[15]$. 

Upon concluding all the iterations, we successfully obtained all the key bytes residing in both LFSR-A ($A[8]$ to $A[15]$) and LFSR-B ($B[8]$ to $B[15]$). Consequently, the incremental attack is effective in recovering all key bytes.

\vspace{-1mm}
\section{SNOW-SCA: Measurement Results}
\subsection{Measurement Setup}

\begin{figure}[!t]
\centerline{\includegraphics[scale=0.18]{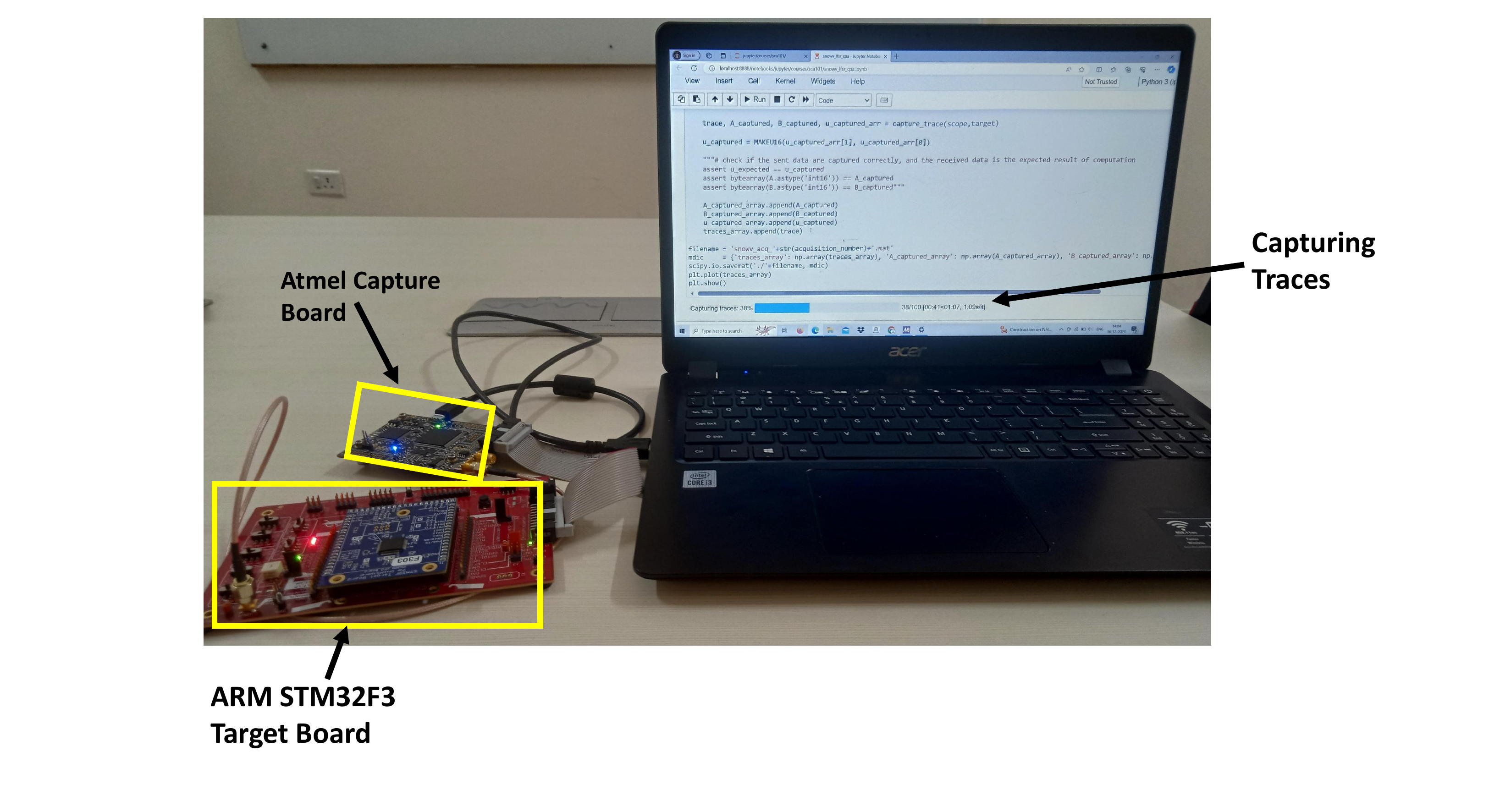}}
\caption{Experimental setup to capture traces using the ChipWhisperer capture board and the target 32-bit ARM Cortex-M4 microcontroller running the SNOW-V algorithm. }
\label{measuring_setup}

\end{figure}

\begin{figure}[!b]
\vspace{-2mm}
\centerline{\includegraphics[scale=0.17]{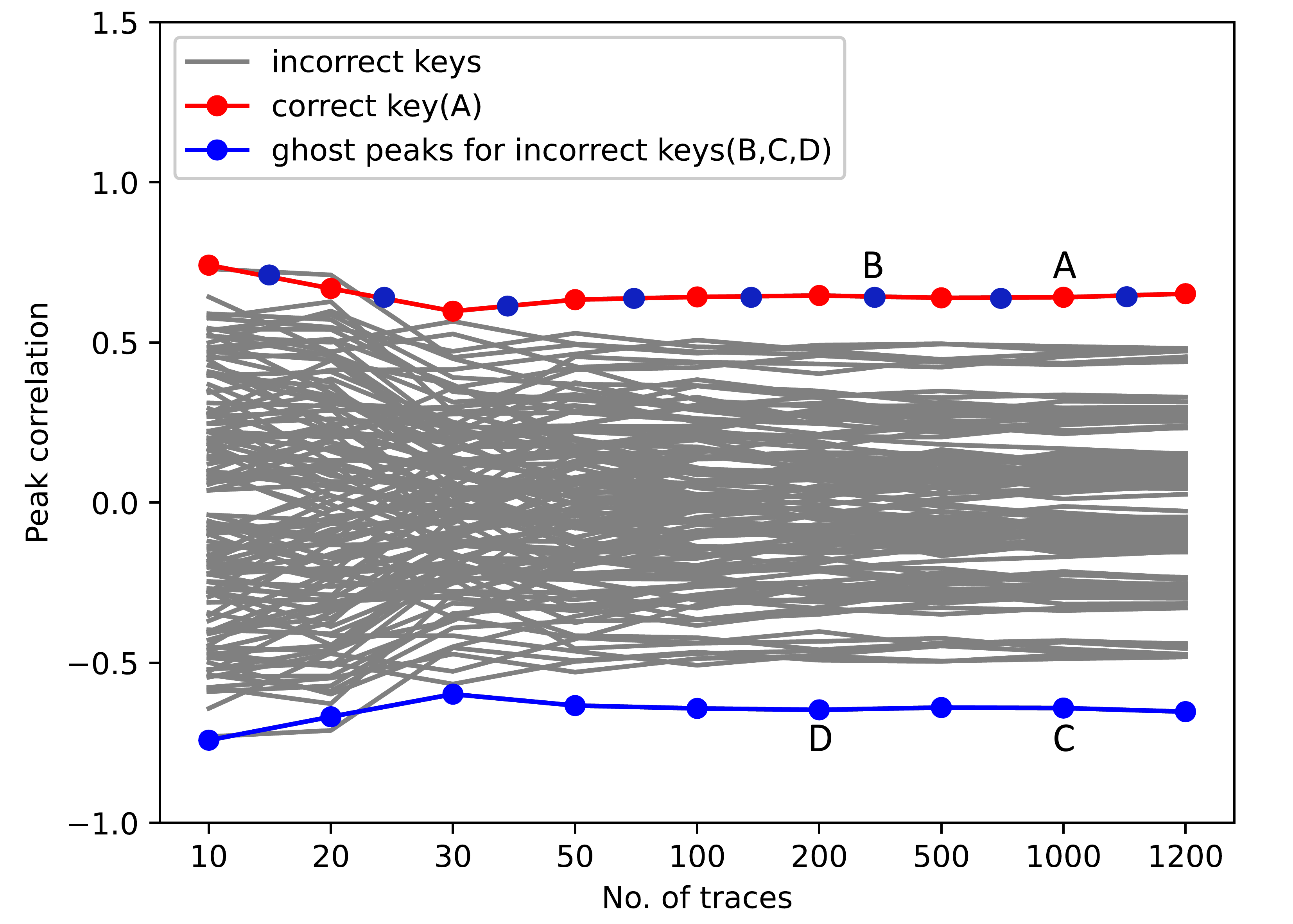}}
\caption{MTD plot (measured): CPA attack on the first key byte for the unprotected SNOW-V implementation.}
\label{mtd without masking}
\end{figure}
\vspace{-1.5mm}
The entire process of capturing power traces for measurements was conducted using the ChipWhisperer platform. Specifically, a ChipWhisperer capture board and a 32-bit ARM STM32F3 target board were employed in this setup, as shown in Fig~\ref{measuring_setup}.

The ChipWhisperer capture board provides a Low-Noise Amplifier (LNA), offering adjustable gain of up to +60 dB, specifically designed for analog power measurements. Equipped with a 10-bit Analog-to-Digital Converter (ADC) capable of reaching up to 105 MS/s, it features an ultra-flexible clocking mechanism that facilitates synchronous power capture, whereas the 32-bit ARM Cortex-M4 STM32F3 is equipped with 40 KB of Static Random-Access Memory (SRAM) and possesses a flash memory capacity of 256 KB. The SNOW-V runs at a frequency of 7.37MHz on the ARM Cortex-M4, and the ChipWhisperer capture board samples the power traces at 30 Mega-Samples/sec.

Before capturing any power traces, the initial step involves constructing a custom firmware designed for communication with the ChipWhisperer. This enables data transmission to and from the microcontroller, facilitating a seamless exchange of information. We plan to open-source the custom firmware for the communication with ChipWhisperer, along with all the captured traces and codes for the CPA and LDA-based attack.

\begin{figure}[!t]
\centerline{\includegraphics[scale=0.16]{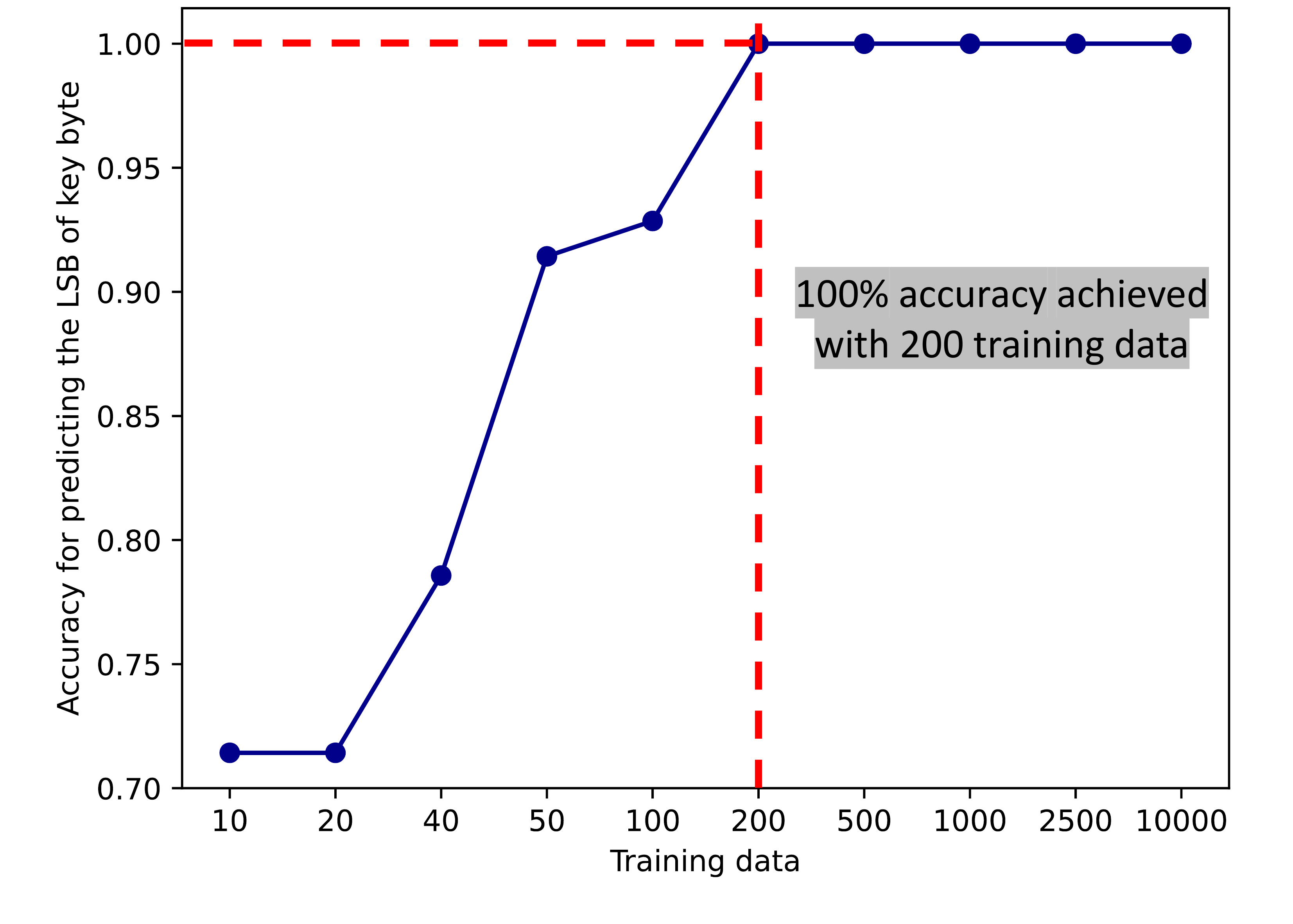}}
\caption{Training accuracy of the LDA model for identifying the LSB of the key byte under attack.}
\vspace{-4mm}
\label{accuracy plot}
\end{figure}
\subsection{SNOW-SCA Attack Results}
The suggested architecture of the SNOW-V stream cipher underwent analysis, and CPA was used to determine the correct key bytes. The results of the CPA exhibited ghost peaks for incorrect keys. This occurrence was due to the shift in the LSB of A[8], as discussed in the previous section. The MTD for CPA on the measured SNOW-V traces shows that the correct key is recovered with $< 50$ traces (Fig.~\ref{mtd without masking}). The MTD plot (Fig.~\ref{mtd without masking}) illustrates that in the case of positive correlation, there is an overlap between one incorrect key (B) and the correct key(A). 

To uniquely identify the correct key byte, we trained an LDA model, which then correctly predicts the LSB of the key byte under attack. The LDA model achieves $100\%$ accuracy after training with 200 traces (Fig.~\ref{accuracy plot}). 

\section{Discussions and Possible countermeasures}~\label{sec:countermeasures}
In this section, we discuss the possible countermeasures to the SNOW-V software implementation to prevent both CPA and LDA-based attacks.

\vspace{-1mm}
\subsection{Constant-time Implementation of the $mul\_x\_inv()$ function}
\vspace{-1.5mm}
As discussed in Section~\ref{sec:analysis_LFSR}, we exploit the leakage from the last bit of $A[8]$ to train our LDA model as explained in Section~\ref{sec:LDA_model}. As shown in the Listing~\ref{lst:non-const}, in the reference implementation~\cite{ekdahl_new_2019}, this portion is implemented in a non-constant time. 
\begin{lstlisting}[caption=Non-constant-time code $mul\_x\_inv$, label={lst:non-const}, language=C]
u16 mul_x_inv(u16 v, u16 d)
{ if (v & 0x0001) return(v >> 1) ^ d;
else return (v >> 1);
}
\end{lstlisting}
However, we observed no reduction in accuracy in our LDA predictor even if we transformed this code to a constant-time code as shown in Listing~\ref{lst:const}. This happens because the LDA model learns from the differences in the power consumption information to train our LDA model instead of the timing information.

\begin{lstlisting}[caption=Constant-time code of $mul\_x\_inv$, label={lst:const}, language=C]
u16 mul_x_inv(u16 v, u16 d)
{ u16 i;
i = v & 0x0001;
return ( i*((v>>1)^d) + (1-i)*(v>>1) )
}
\end{lstlisting}

\subsection{Boolean masking on the points of attack in $lfsr\_update()$ }
Masking~\cite{ChaJutRaoRoh1999} is a provable secure countermeasure that protects cryptographic implementations against passive side-channel attacks, including CPA. There are several kinds of masking techniques, such as Boolean masking~\cite{DBLP:conf/ches/RivainP10}, multiplicative masking~\cite{DBLP:conf/ches/RivainP10}, additive masking~\cite{conf/ches/CoronG00}, etc. These masking techniques are integrated with cryptographic operations to construct a secure implementation efficiently, depending on the type of operations. SNOW-V mainly uses Boolean operations, so it would be efficient to integrate Boolean masking to secure SNOW-V implementation against SCA. In Boolean masking, we divide any sensitive variable $x$ (a variable that is generated from an interaction with the secret key) into multiple shares (for first-order SCA protection, we split $x$ into two shares $x_1$ and $x_2$, such that $x_1 \oplus x_2 = x$) and then perform all the operations of the cryptographic algorithm separately on these two shares. 

\begin{figure}[!t]
\centerline{\includegraphics[scale=0.15]{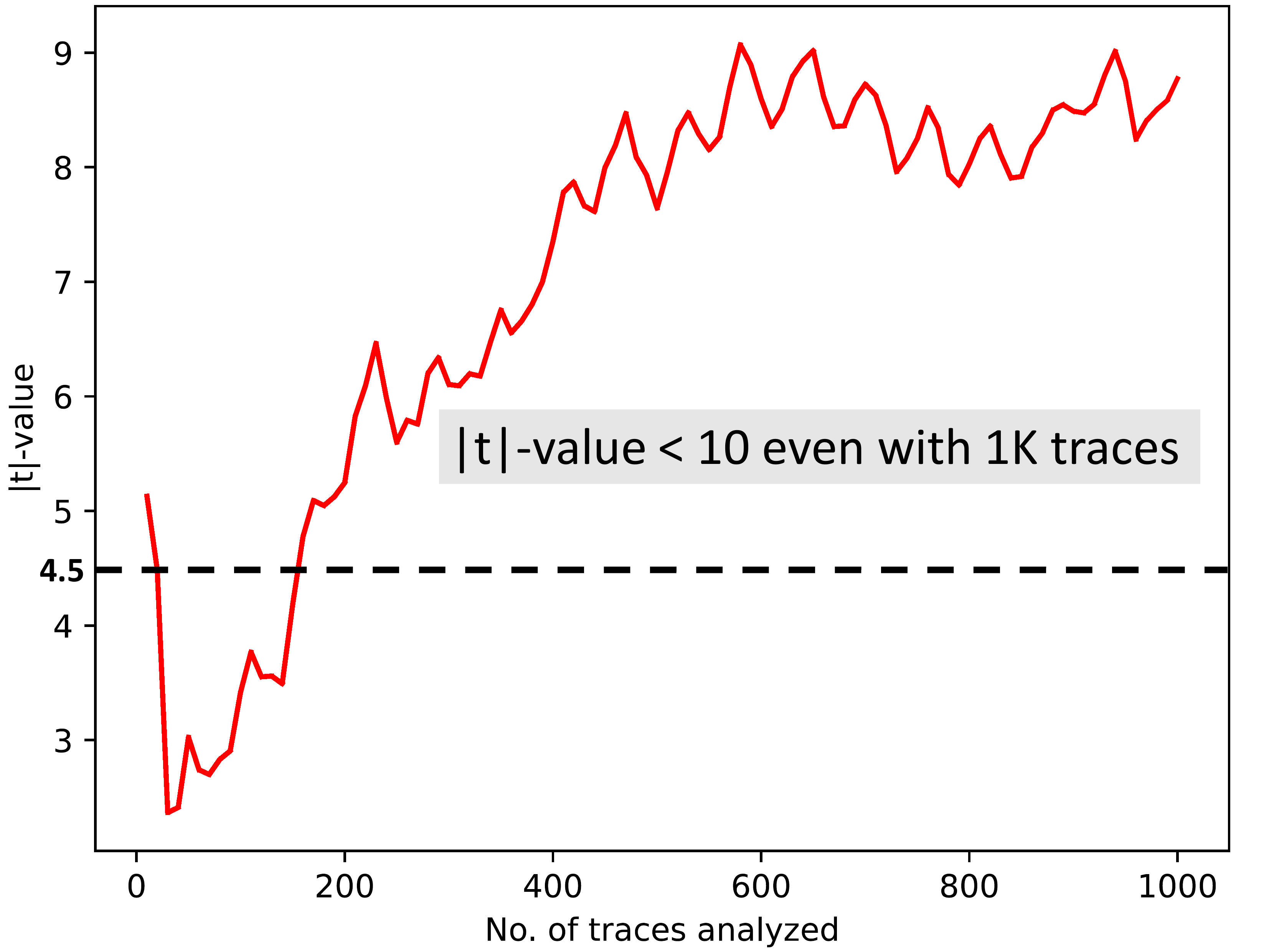}}
\caption{Fixed-vs-random TVLA shows that the $|t|$-value remains $<10$ for 1K traces (1K for each set - fixed and random IV). }
\label{tvla_with_masking}
\end{figure}



Since our attack target here is the $u$ and $v$ variables in the $lfsr\_update()$, which comprises the Boolean operations, we use a Boolean masking scheme. For the first-order masking, we utilize a 16-bit random variable to mask the values of $u$ and $v$ as $u \oplus r$ and $v \oplus r$, respectively. This will prevent revealing the hamming weights of the variables $u$ and $v$ to an attacker with a single probe. 


The masked SNOW-V implementation is validated using the ChipWhisperer measurement framework (Fig. \ref{measuring_setup}). As shown in Fig. \ref{tvla_with_masking}, the fixed-vs-random TVLA test on the masked SNOW-V implementation reveals that the $|t|$-value remains $<10$ even with 1K analyzed traces (1K from each set - random and fixed IV), compared to $\sim 150$ for the unprotected SNOW-V implementation with the same number of traces (refer to Fig. \ref{TVLA}(b)). 

\begin{figure}[!t]
\centerline{\includegraphics[scale=0.17]{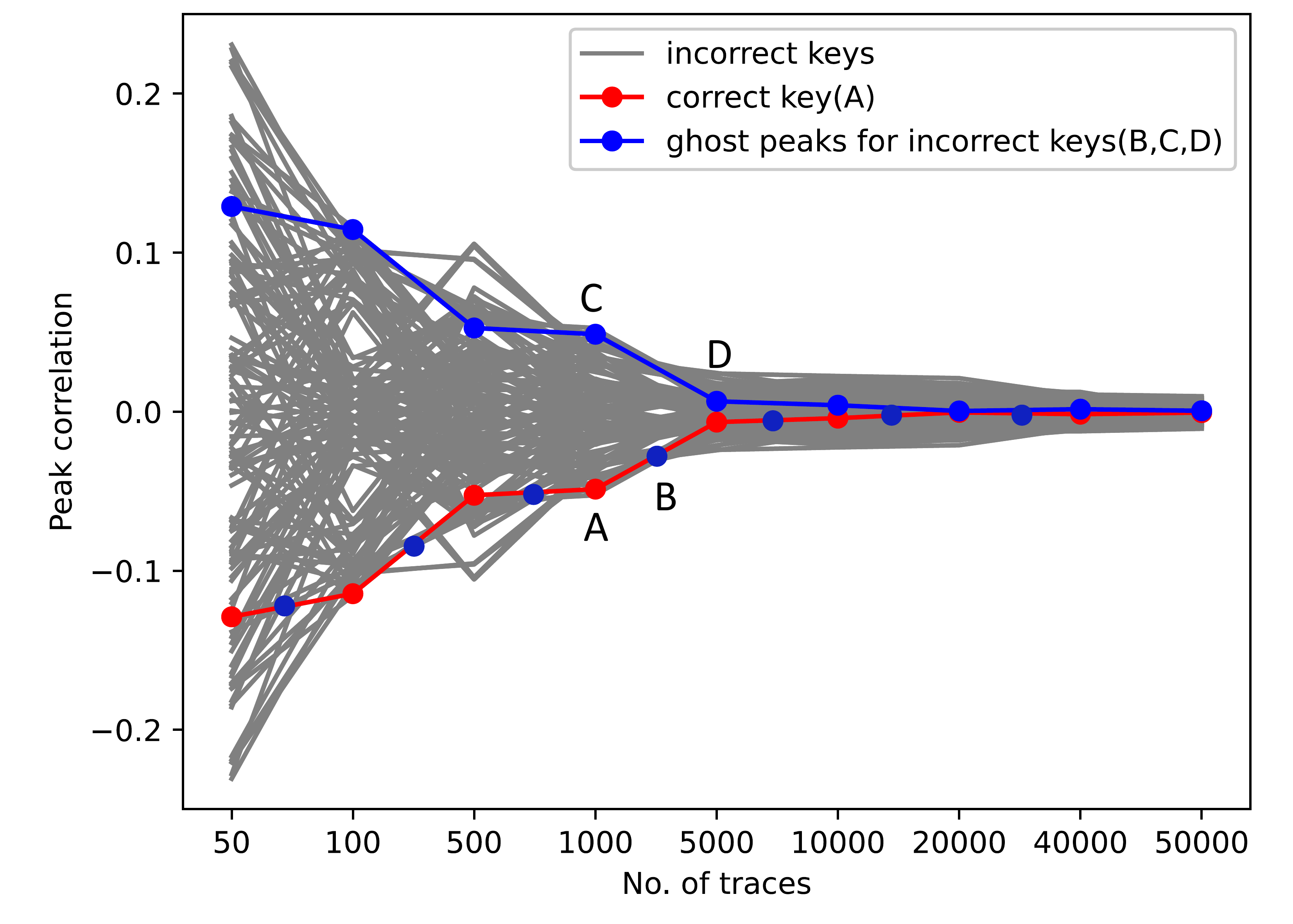}}
\caption{Measured CPA results on the masked SNOW-V implementation shows that the secret key byte could not be extracted even after $50K$ traces. Hence, the proposed boolean masking technique provides an SCA improvement of $>1000\times$ against our proposed SNOW-SCA attack. }
\label{mtd_with_mask}
\end{figure}

Fig. \ref{mtd_with_mask} shows the measured CPA results on the masked SNOW-V implementation. We can clearly see that the correct key could not be revealed even after $50K$ traces, showing a $>1000\times$ SCA security improvement. This highlights the efficacy of the proposed countermeasure using Boolean masking. However, since the $|t|$-value crosses the threshold of 4.5, there exists some data-dependent leakage, which may be exploited by attacking other points of the SNOW-V implementation. Hence, we must develop a full-fledged masking scheme for the entire algorithm (instead of just the $u$ and $v$ attack points in the $lfsr\_update()$). However, it has consequences in terms of throughput degradation. We discuss this in more detail in Section V-D.

\subsection{Shuffling of the LFSR Rounds}
Alternatively, a low-cost countermeasure on the algorithm is shuffling~\cite{DBLP:conf/acns/HerbstOM06}, where the execution order of the independent operation gets shuffled (using a permutation function) and makes the processor non-deterministic. Here, although the operations generate power leakages depending on the processing data, the time when the operations are performed varies every execution. Shuffling techniques increase the complexity of the side-channel trace requirements to perform SCA in the presence of enough noise in the channel. The attack complexity increases linearly with the number of independent operations on which the permutation is applied. In the case of SNOW-V, the first five iterations within the $lfsr\_update()$ function are independent and can be shuffled. However, it is challenging to apply the shuffling technique on the entire $lfsr\_update()$ function, as the last three iterations consist of dependent operations and conversion of most of the operations of $lfsr\_update()$ to a set of independent operations efficiently is challenging and needs more research. We also want to note that there is little research on side-channel attacks and their countermeasures on stream ciphers~\cite{survey_SCA_countermeasure}.




\subsection{Future Directions}
For masked implementations, we are repeating the operations on different shares independently. It significantly increases ($>2\times$) the run time of the whole cryptographic algorithm. We can do this parallelly on hardware platforms with more area and power consumption. Since the 5G ciphers have very stringent limits on performance and resource consumption, developing suitable masking schemes is challenging. 

On the other hand, recently, various low-overhead circuit-level countermeasures like switched capacitor current equalizer \cite{tokunaga_secure_2009}, integrated voltage regulators (IVR) \cite{singh_25.3_2019}, and signature attenuation using STELLAR \cite{das_stellar_2019, das_high_2017, das_273_2020, ghosh_synstellar_2022, das_asni:_2018} have been presented. While logical countermeasures such as masking suffer from high overheads and performance degradation and are algorithm-specific, circuit-level countermeasures are algorithm-agnostic and typically have the lowest overheads for the same level of SCA security \cite{das_273_2020, das_em_2020, ghosh_362_2021}. Such circuit-level countermeasures can be used as a generic wrapper around the microcontroller core, ensuring an almost constant supply current to an external attacker and thus providing resilience to any crypto or security-sensitive algorithm running in software.

Finally, it should be worth noting that developing an efficient countermeasure for the full encryption (or decryption) of a cryptographic algorithm is a non-trivial task that requires a significant amount of research and experimentation. Throughout this section, we have laid down some roadmap and possible directions that could be beneficial in developing a suitable countermeasure. Our initial first-order masking experiments also corroborate some of these hypotheses. Developing a full-fledged SCA-protected implementation of SNOW-V is part of our future research.

\section{Conclusion}
In summary, this paper presents SNOW-SCA, the first power SCA attack on the 5G standard candidate SNOW-V. Utilizing a combined CPA and LDA attack, the full secret key is recovered for the software implementation of SNOW-V running on a 32-bit ARM Cortex-M4 microcontroller. While the CPA narrows down the hypothesis to two key byte guesses, the LSB for the byte under attack could not be determined uniquely as it gets thrown away by an intermediate operation ($mul\_x\_inv()$) within the $lfsr\_update()$. Hence, an LDA model is trained to predict the LSB based on the branching condition in the $mul\_x\_inv()$ function. LDA achieves $100\%$ accuracy with $<200$ training traces. Overall, the correct key byte is recovered in $<50$ traces using our proposed attack strategy. Finally, we demonstrated boolean masking to specifically protect the points of attack, which shows successful SCA resilience even with $50K$ traces.

\bibliographystyle{IEEEtran}
{
\bibliography{main}
}

\end{document}